\documentstyle[12pt,epsf]{article}
\newlength{\pubnumber} 

\catcode`\@=11
\@addtoreset{equation}{section}

\def\section{\@startsection{section}{1}{\z@}{3.5ex plus 1ex minus .2ex}
 {2.3ex plus .2ex}{\large\bf}}
\def\subsection{\@startsection{subsection}{2}{\z@}{2.3ex plus .2ex}
 {2.3ex plus .2ex}{\bf}}

\begin{document}
\begin{titlepage}
\samepage{
\setcounter{page}{1}
\rightline{IASSNS-HEP-96/100}
\rightline{\tt hep-ph/9610479}
\rightline{published in {\it Nucl.\ Phys.}\/ {\bf B492} (1997) 104}
\rightline{October 1996}
\vfill
\vfill
\begin{center}
 {\Large \bf Kinetic Mixing and the Supersymmetric\\ Gauge
     Hierarchy \\}
\vfill
\vfill
 {\large
   Keith R. Dienes,\footnote{E-mail address: dienes@sns.ias.edu}$\,$
   Christopher Kolda,\footnote{E-mail address: kolda@sns.ias.edu}
   $\,$and$\,$
   John March-Russell\footnote{E-mail address: jmr@sns.ias.edu}\\
  }
\vspace{.25in}
 {\it  School of Natural Sciences, Institute for Advanced Study\\
  Olden Lane, Princeton, N.J.~~08540~ USA\\}
\end{center}
\vfill
\vfill
\begin{abstract}
  {\rm
	The most general Lagrangian for a model with two
	$U(1)$ gauge symmetries contains a renormalizable operator
	which mixes their gauge kinetic
	terms. Such kinetic mixing can be generated at
	arbitrarily high scales but will not be suppressed by large masses.
	In models whose supersymmetry (SUSY)-breaking hidden
	sectors contain $U(1)$ gauge factors, we show that such terms will
	generically arise and communicate SUSY-breaking to the visible
	sector through mixing with hypercharge.
	In the context of the usual
	supergravity- or gauge-mediated communication scenarios with
	$D$-terms of order the fundamental scale of SUSY-breaking,
	this effect can destabilize the gauge hierarchy.
	Even in models for which kinetic
	mixing is suppressed or the $D$-terms are arranged to be small,
	this effect is a potentially large
	correction to the soft scalar masses and
        therefore introduces a new measurable low-energy parameter.
	We calculate the size of kinetic mixing both in field theory and
	in string theory, and argue that appreciable kinetic mixing
	is a generic feature of string models.  We conclude that the
	possibility of kinetic mixing effects cannot be ignored in
	model-building and in phenomenological studies of the low-energy
	SUSY spectra.
   }
\end{abstract}
\vfill}
\end{titlepage}

%======================= DEFINITIONS =====================================

\catcode`@=11
% Redefine caption to put text and formulas in smaller font
\long\def\@caption#1[#2]#3{\par\addcontentsline{\csname
  ext@#1\endcsname}{#1}{\protect\numberline{\csname
  the#1\endcsname}{\ignorespaces #2}}\begingroup
    \small
    \@parboxrestore
    \@makecaption{\csname fnum@#1\endcsname}{\ignorespaces #3}\par
  \endgroup}
\catcode`@=12

\newcommand{\newc}{\newcommand}
\newc{\gsim}{\lower.7ex\hbox{$\;\stackrel{\textstyle>}{\sim}\;$}}
\newc{\lsim}{\lower.7ex\hbox{$\;\stackrel{\textstyle<}{\sim}\;$}}
\newc{\gev}{\,{\rm GeV}}
\newc{\mev}{\,{\rm MeV}}
\newc{\ev}{\,{\rm eV}}
\newc{\kev}{\,{\rm keV}}
\newc{\tev}{\,{\rm TeV}}
\def\tr{\mathop{\rm tr}}
\def\Tr{\mathop{\rm Tr}}
\def\Im{\mathop{\rm Im}}
\def\Re{\mathop{\rm Re}}
\def\bR{\mathop{\bf R}}
\def\bC{\mathop{\bf C}}
\def\lie{\mathop{\hbox{\it\$}}} %pound sterling
\newc{\sw}{s_W}
\newc{\cw}{c_W}
\newc{\swsq}{s^2_W}
\newc{\swsqb}{s^2_W}
\newc{\cwsq}{c^2_W}
\newc{\cwsqb}{c^2_W}
\newc{\Qeff}{Q_{\rm eff}}
\newc{\mz}{M_Z}
\newc{\mpl}{M_{\rm Pl}}
\renewcommand{\phi}{\varphi}
%
%%%%%%%%%%%%%%%%%% Reference Defs %%%%%%%%%%%%%%%%%%
%
\def\NPB#1#2#3{{\it Nucl.\ Phys.}\/ {\bf B#1} (19#2) #3}
\def\PLB#1#2#3{{\it Phys.\ Lett.}\/ {\bf B#1} (19#2) #3}
\def\PLBold#1#2#3{{\it Phys.\ Lett.}\/ {\bf B#1} (19#2) #3}
\def\PRD#1#2#3{{\it Phys.\ Rev.}\/ {\bf D#1} (19#2) #3}
\def\PRL#1#2#3{{\it Phys.\ Rev.\ Lett.}\/ {\bf#1} (19#2) #3}
\def\PRT#1#2#3{{\it Phys.\ Rep.}\/ {\bf#1} (19#2) #3}
\def\ARAA#1#2#3{{\it Ann.\  Rev.\  Astron.\  Astrophys.}\/ {\bf#1} (19#2) #3}
\def\ARNP#1#2#3{{\it Ann.\ Rev.\ Nucl.\ Part.\ Sci.}\/ {\bf#1} (19#2) #3}
\def\MPL#1#2#3{{\it Mod.\ Phys.\ Lett.}\/ {\bf #1} (19#2) #3}
\def\ZPC#1#2#3{{\it Zeit.\ f\"ur\ Physik}\/ {\bf C#1} (19#2) #3}
\def\APJ#1#2#3{{\it Ap.\ J.}\/ {\bf #1} (19#2) #3}
\def\AP#1#2#3{{\it Ann.\ Phys.}\/ {\bf #1} (19#2) #3}
\def\RMP#1#2#3{{\it Rev.\ Mod.\ Phys.}\/ {\bf #1} (19#2) #3}
\def\CMP#1#2#3{{\it Comm.\ Math.\ Phys.}\/ {\bf #1} (19#2) #3}
\def\IJMP#1#2#3{{\it Int.\ J.\ Mod.\ Phys.}\/ {\bf A#1} (19#2) #3}
\relax
%
%
%%%%%%%%%%%%%%%%%%%%%%% latex eqn abrev's %%%%%%%%%%%%%%%%%%%%%%%%%%%%
%
\def\beq{\begin{equation}}
\def\eeq{\end{equation}}
\def\bea{\begin{eqnarray}}
\def\eea{\end{eqnarray}}
%
%%%%%%%%%%%%%%%%%%% special features for eqns %%%%%%%%%%%%%%%%
%
%     this boxes an equation
%
\def\boxeqn#1{\vcenter{\vbox{\hrule\hbox{\vrule\kern3pt\vbox{\kern3pt
\hbox{${\displaystyle #1}$}\kern3pt}\kern3pt\vrule}\hrule}}}
%
%     this draws a little box (end of proof symbol)
%     e.g. \qed{.1}{.1}
%
\def\qed#1#2{\vcenter{\hrule \hbox{\vrule height#2in
\kern#1in \vrule} \hrule}}
\def\half{{\textstyle{1\over2}}} %%small half in a displayed eqn
%\def\frac#1#2{{\textstyle{#1\over #2}}} %%small fraction in a displayed eqn
%
%
%%%%%%%%%%%%%%%%%%%%%%% common abrev's %%%%%%%%%%%%%%%%%
%
%
\newc{\ie}{{\it i.e.}\/}          \newc{\etal}{{\it et al.}\/}
\newc{\eg}{{\it e.g.}\/}          \newc{\etc}{{\it etc.}\/}
\newc{\cf}{{\it c.f.}\/}
%
%
%%%%%%%%%%%%%%%%%%%%%%%% curly letters %%%%%%%%%%%%%%%%%%%
%
%
\def\CAG{{\cal A/\cal G}}
\def\CA{{\cal A}} \def\CB{{\cal B}} \def\CC{{\cal C}} \def\CD{{\cal D}}
\def\CE{{\cal E}} \def\CF{{\cal F}} \def\CG{{\cal G}} \def\CH{{\cal H}}
\def\CI{{\cal I}} \def\CJ{{\cal J}} \def\CK{{\cal K}} \def\CL{{\cal L}}
\def\CM{{\cal M}} \def\CN{{\cal N}} \def\CO{{\cal O}} \def\CP{{\cal P}}
\def\CQ{{\cal Q}} \def\CR{{\cal R}} \def\CS{{\cal S}} \def\CT{{\cal T}}
\def\CU{{\cal U}} \def\CV{{\cal V}} \def\CW{{\cal W}} \def\CX{{\cal X}}
\def\CY{{\cal Y}} \def\CZ{{\cal Z}}
%
%
%
%%%%%%%%%%%%%%%%%%% derivatives %%%%%%%%%%%%%%%%%%%%%%%%%%%%%
%
%
\def\grad#1{\,\nabla\!_{{#1}}\,}
\def\gradgrad#1#2{\,\nabla\!_{{#1}}\nabla\!_{{#2}}\,}
\def\partder#1#2{{\partial #1\over\partial #2}}
\def\secder#1#2#3{{\partial^2 #1\over\partial #2 \partial #3}}
%
%
%
%%%%%%%%%%%%%%%%%%%%% relations %%%%%%%%%%%%%%%%%%%%%%%%%%%%%
%
%
\def\ltap{\ \raise.3ex\hbox{$<$\kern-.75em\lower1ex\hbox{$\sim$}}\ }
\def\gtap{\ \raise.3ex\hbox{$>$\kern-.75em\lower1ex\hbox{$\sim$}}\ }
\def\gl{\ \raise.5ex\hbox{$>$}\kern-.8em\lower.5ex\hbox{$<$}\ }
\def\roughly#1{\raise.3ex\hbox{$#1$\kern-.75em\lower1ex\hbox{$\sim$}}}
%
%
%%%%%%%%%%%%%%%%%%%% slashed symbols %%%%%%%%%%%%%%%%%%%%%
%
%
\def\slash#1{\rlap{$#1$}/} % slashes a character
\def\dsl{\,\raise.15ex\hbox{/}\mkern-13.5mu D} %this one can be subscripted
\def\delsl{\raise.15ex\hbox{/}\kern-.57em\partial}
\def\Ksl{\hbox{/\kern-.6000em\rm K}}
\def\Asl{\hbox{/\kern-.6500em \rm A}}
\def\Dsl{\hbox{/\kern-.6000em\rm D}} %roman D
\def\Qsl{\hbox{/\kern-.6000em\rm Q}}
\def\gradsl{\hbox{/\kern-.6500em$\nabla$}}
%
%%%%%%%%%%%%%%%%%%% greek letters %%%%%%%%%%%%%%%%%%%%
%
\let\al=\alpha
\let\be=\beta
\let\ga=\gamma
\let\Ga=\Gamma
\let\de=\delta
\let\De=\Delta
\let\ep=\varepsilon
\let\ze=\zeta
\let\ka=\kappa
\let\la=\lambda
\let\La=\Lambda
\let\del=\nabla
\let\si=\sigma
\let\Si=\Sigma
\let\th=\theta
\let\Up=\Upsilon
\let\om=\omega
\let\Om=\Omega
\def\ph{\varphi}
%
%%%%%%%%%%%%%%%%% BOLD greek letters %%%%%%%%%%%%%%%%%%%%%%%%%%%%%%
%
% Style-sensitive Poor-Man's-Bold command, produces bold greek letters.
% Usage $ ... \pmb\gamma ... $
% Adapted from TeXbook p386 (\pmb) and p360 (\mathpallette)
%
\newdimen\pmboffset
\pmboffset 0.022em
\def\oldpmb#1{\setbox0=\hbox{#1}%
 \copy0\kern-\wd0
 \kern\pmboffset\raise 1.732\pmboffset\copy0\kern-\wd0
 \kern\pmboffset\box0}
\def\pmb#1{\mathchoice{\oldpmb{$\displaystyle#1$}}{\oldpmb{$\textstyle#1$}}
	{\oldpmb{$\scriptstyle#1$}}{\oldpmb{$\scriptscriptstyle#1$}}}
%
%
%
%%%%%%%%%%%%%%%%%%% various symbol abbreviations, vev's etc %%%%%%%%%%%
%
%
\def\bar#1{\overline{#1}}
\def\vev#1{\left\langle #1 \right\rangle}
\def\bra#1{\left\langle #1\right|}
\def\ket#1{\left| #1\right\rangle}
\def\abs#1{\left| #1\right|}
\def\vector#1{{\vec{#1}}}
\def\inv{^{\raise.15ex\hbox{${\scriptscriptstyle -}$}\kern-.05em 1}}
\def\pr#1{#1^\prime}  %prime
\def\lbar{{\lower.35ex\hbox{$\mathchar'26$}\mkern-10mu\lambda}} %lambda bar
\def\e#1{{\rm e}^{^{\textstyle#1}}}
\def\ee#1{\times 10^{#1} }
\def\om#1#2{\omega^{#1}{}_{#2}}
\def\imp{~\Rightarrow}
\def\coker{\mathop{\rm coker}}
\let\p=\partial
\let\<=\langle
\let\>=\rangle
\let\ad=\dagger
\let\txt=\textstyle
\let\h=\hbox
\let\+=\uparrow
\let\-=\downarrow
\def\dot{\!\cdot\!}
\def\vfilll{\vskip 0pt plus 1filll}
%
%%%%%%%%%%%%%%%%%%% end of defns %%%%%%%%%%%%%%%%%%%%%%%%%%%%%%%%%%%%%

\setcounter{footnote}{0}
\setcounter{section}{0}
\setcounter{subsection}{0}
\setcounter{subsubsection}{0}

%%%%%%%%%%%%%%%%%%%%%%%%%%%%%%%%%%%%%%%%%%%%%%%%%%%%%%%%%%%%%%%%%%%%%%%

\setcounter{footnote}{0}
\section{Introduction} \label{sec:intro}

Modern models of supersymmetry (SUSY)-breaking in the minimal
supersymmetric Standard Model (MSSM)
always involve the division of the full theory into a
so-called ``hidden'' sector and the usual ``visible'', or MSSM, sector.
This avoids the problem that arises if SUSY is broken
in a sector with tree-level couplings to the MSSM, namely the
existence of experimentally excluded sum rules on the MSSM sparticle
masses~\cite{sumrules}. Given the reasonable expectation that SUSY should
be broken by non-trivial dynamics in the infrared, the hidden sector
must contain a non-abelian gauge group $G$, and in this case ``hidden'' implies
that there exist no tree-level interactions which couple states charged
under $G$ with those charged under the MSSM. In particular,
there must not exist fields in the effective Lagrangian (below the scale
of SUSY-breaking) which are charged under both gauge groups simultaneously.

Given that SUSY-breaking is generated by hidden-sector dynamics,
the most pertinent issue for MSSM phenomenology is the nature of
the communication mechanism from the hidden to MSSM sector.
The simplest known mechanism is supergravity (SUGRA),
which couples the hidden to the visible sector through
operators which are Planck-scale suppressed. One assumes that SUSY is broken
at some high scale (typically $\Lambda^2\sim\mz\mpl$), so that the
SUSY-breaking mass scale which is
communicated to the visible sector by SUGRA is
$\Lambda^2/\mpl\sim\mz$~\cite{sugra}.

Recently another class of models has received
attention. These models communicate SUSY-breaking through a cascade of gauge
interactions, some of which are those of the MSSM and some those of
the hidden sector.  Here the SUSY-breaking scale communicated to
the visible sector is suppressed by loop
factors compared to the scale of SUSY-breaking in the hidden sector,
typically taken to be $\sim 100\tev$~\cite{oldgmsb,newgmsb,dnns}.

In either case, it is imperative for the consistency of
the model that there not exist operators which couple the larger
hidden-sector SUSY-breaking scale to the visible sector without the appropriate
loop or $\mpl$ suppressions.  Such an operator would pull the weak scale up
to the SUSY-breaking scale in the hidden sector, thereby
destabilizing the gauge hierarchy.

It has been known for some time that there can exist operators
in the effective Lagrangian which perform just such an unwanted task. In the
cases usually studied, such operators have the form of tadpoles. Since
tadpoles of chiral superfields involve only gauge singlets, it has been
noted that the existence of such singlets can have potentially disastrous
consequences~\cite{bagger}.

In this paper we will consider a new communication mechanism for
SUSY-breaking which has the potential to destabilize any model
with a $U(1)$ gauge factor in the hidden sector whose $D$-terms are
of order the fundamental SUSY-breaking scale $\La^2$.
This communication is provided by a renormalizable operator
which is often overlooked, namely the mixing of two
separate $U(1)$ gauge kinetic functions.
Because this operator is renormalizable, it may be
generated by physics at scales far above the SUSY-breaking scale itself,
without any suppression by the large mass scale.  Furthermore,
we will argue that this operator is generated at one loop in generic
field theory models,
and is expected to occur quite naturally in realistic string models.
It can therefore easily dominate over both SUGRA-mediated
and gauge-mediated (GM) soft scalar mass terms.
Even in models for which this not the case, it is still quite
possible that this new contribution significantly corrects the
usual soft scalar masses generated by SUGRA or GM without destroying
the gauge hierarchy.

This paper is organized as follows.
In Sect.~2, we discuss the consequences
of kinetic mixing for supersymmetric theories.
In Sect.~3, we then discuss
the sources of kinetic mixing, show how
to calculate kinetic mixing effects in string theory,
and estimate typical sizes that may be expected within the context of both
field theory and string theory.
Finally, in Sect.~4, we present our conclusions.

\setcounter{footnote}{0}
\section{Supersymmetric Kinetic Mixing}

It was realized many years ago~\cite{holdom}\ that in
a theory with two $U(1)$ gauge
factors, there can appear in the Lagrangian a term
which is consistent with all gauge symmetries and which mixes the two $U(1)$'s.
In the basis in which  the interaction terms have the canonical form, the
pure gauge part of the Lagrangian for an arbitrary $U(1)_a\times U(1)_b$
theory can be written
\beq
\CL_{\rm gauge}~=~ -\frac{1}{4}\,F_{(a)}^{\mu\nu}F_{(a)\mu\nu}
-\frac{1}{4}\,F_{(b)}^{\mu\nu}F_{(b)\mu\nu}
+\frac{\chi}{2}\,F_{(a)}^{\mu\nu}F_{(b)\mu\nu}~.
\label{firstequation}
\eeq
(Throughout this analysis we will work to leading order in
$\chi$ for simplicity.)
In a supersymmetric theory, such a
Lagrangian generalizes to\footnote{
      After completing this work we became aware
      of Ref.~\cite{pol}\ in which the existence of
      the operator $W_aW_b$ was briefly mentioned.}
\beq
  \CL_{\rm gauge}~=~ \frac{1}{32}\int d^2\th\,\left\{W_aW_a+W_bW_b-2\chi W_aW_b
\right\}
\eeq
where $W_a$ and $W_b$ are the chiral gauge field strength superfields for
the two gauge symmetries: $W=\bar D^2DV$ for the vector superfield $V$.
In principle, both $U(1)$'s could lie in the hidden
sector, or both in the visible, but we will primarily interest ourselves with
the case in which
$U(1)_a$ is in the visible sector while
$U(1)_b$ is in the SUSY-breaking hidden sector.
Hypercharge is an example of such a $U(1)_a$.

To bring the pure gauge portion of the Lagrangian to canonical form, one
can shift the visible-sector gauge field:
\beq
V^\mu_a~\to~ V'^\mu_a=V^\mu_a - \chi V^\mu_b
\eeq
which implies that $W_a\to W'_a=W_a-\chi W_b$. This particular choice of
basis is dictated by the assumption that $U(1)_b$ will break once
some hidden-sector field with non-zero charge receives a vacuum expectation
value.
In this basis, the gauge Lagrangian is diagonal:
\beq
\CL_{\rm gauge}~=~\frac{1}{32}\int d^2\th\,\left\{W'_aW'_a+ W_bW_b \right\}~.
\eeq
However, the same shift must also be performed in the interaction piece:
\bea
  \CL_{\rm int}&=&\int d^4\th\,\left\{
               \phi_i^\dagger e^{2g_aV_a}\phi_i
              +\Phi_i^\dagger e^{2g_bV_b}\Phi_i
                   \right\} \nonumber \\
       &=&\int d^4\th\,\left\{
         \phi_i^\dagger e^{2g_aV'_a+2g_a\chi V_b}\phi_i
        +\Phi_i^\dagger e^{2g_bV_b}\Phi_i
                 \right\}
\label{eq:intshift}
\eea
where the final term is in the basis in which the kinetic terms are
diagonal.  Here we denote by
$\phi_i$ those chiral superfields charged under only $U(1)_a$ (the visible
sector),
and by $\Phi_i$ those charged under only $U(1)_b$ (the hidden sector).

Eq.~(\ref{eq:intshift}) implies a number of new interactions between
the visible and hidden sectors. First, the visible-sector states
obtain hidden charges proportional to
their visible-sector charges and couplings:
\beq
    \CD_b^\mu~=~\partial^\mu + i(g_b Q_b + g_a\chi Q_a) V_b^\mu~.
\eeq
Second, the visible-sector fields and their superpartners now couple to the
gauginos of the hidden sector:
\beq
       \CL~=~i\sqrt{2}\,g_aQ_a\chi\phi_i^\dagger \tilde\phi_i\lambda_b ~+~ {\rm
h.c.}
\eeq
Third and finally, upon solving for the $D$-terms via
their equations of motion, one finds:
\bea
     D'_a&=&-g_a\sum_i Q_{ai}|\phi_i|^2~\nonumber\\
     D_b&=&-g_b\sum_i Q_{bi}|\Phi_i|^2~-~\chi\, g_a\sum_i Q_{ai}|\phi_i|^2 ~.
\label{eq:dterms}
\eea
The scalar potential is then given by $V= \frac{1}{2}D'_aD'_a + \frac{1}{2}
D_bD_b$.
Also note that
in the presence of a Fayet-Iliopoulos term $\xi\int d^4\th\,V$ for
$U(1)_b$, we have $D_b\to D_b+\xi$.

Several points are immediately apparent. First, SUSY-breaking has been
communicated to the visible sector via $D_b$, and there arise
new kinetic mixing contributions to the soft masses of visible-sector scalars:
\beq
       (m^2_i)_{\rm KM} ~=~ g_a\,\chi\, Q_{ai} \vev{D_b}~.
\label{eq:mass}
\eeq
These contributions are in addition to any other induced soft masses, \eg:
\beq
   m^2_i~=~(m^2_i)_{\rm GM} + (m^2_i)_{\rm SUGRA} + (m^2_i)_{\rm KM}
       +  g_a Q_{ai}\vev{D_a}~,
\label{eq:allmasses}
\eeq
where the last term is the usual visible-sector $D$-term. In the MSSM,
we have
\beq
g_aQ_{ai}\vev{D_a}~=~ \mz^2\cos2\beta\left(Q_i\cos^2\theta_W - Y_i\right)
\eeq
where $Q$ is ordinary electric charge, $Y$ is the hypercharge, and
$\tan\beta=\vev{H_u}/\vev{H_d}$. A reorganization
of Eq.~(\ref{eq:allmasses}) then leads to
\bea
    (m^2_i)_{\rm
KM}+g_aQ_{ai}\vev{D_a}&=&g_aQ_{ai}\left(\vev{D_a}+\chi\vev{D_b}
      \right) \nonumber \\
      &=&\mz^2\cos2\beta\left[Q_i\cos^2\theta_W - Y_i\left(1+
        \frac{\eta}{\cos2\beta}\right)\right]
\label{eq:x}
\eea
where the final equality holds in the MSSM, and where we have defined
\beq
         \eta~\equiv~\frac{\chi\vev{D_b}}{\mz^2}~.
\eeq
Thus $\eta$ is the parameter which signals the importance of
kinetic mixing effects.

There are three distinct cases to consider, depending on whether
in Eq.~(\ref{eq:x}) we have $\eta\gg1$, $\eta\sim1$, or $\eta\ll1$.
If $\eta\gg1$, then kinetic
mixing is the dominant messenger of SUSY-breaking, leading to two related
disasters.  First, the hierarchy is destabilized, with the scalar masses being
pulled far above $\mz$ to the hidden-sector SUSY-breaking scale.
Second, since $U(1)_a$ anomaly cancellation requires that both signs
of $U(1)_a$ charge exist,
at least one state must receive negative
mass-squared, thereby breaking the symmetries under which it transforms.
For the case of hypercharge, this is obvious, since the states of the MSSM
carry both signs of $Y$.
Thus kinetic mixing with $Y$ cannot be the primary means
for communicating SUSY-breaking to the visible world.

If $\eta\sim1$, then kinetic mixing provides a measurable correction to the
soft scalar masses. It is noteworthy that these corrections are
flavor-independent and do not directly produce additional flavor-changing
neutral current (FCNC) effects.
Extracting such contributions from the usual SUGRA and/or GM soft
masses is likely to be quite involved.  For example, there are no
simple soft mass sum rules which directly measure $\eta$
without reference to the
original ($\eta=0$) soft masses.  Typical relations are of the form:
\beq
m^2_{\tilde e_L}+m^2_{\tilde\nu}~=~2(m^2_{\tilde L})_{\eta=0}
-{\textstyle{1\over 2}}
\mz^2(\cos2\beta+\eta)
\eeq
where $(m^2_{\tilde L})_{\eta=0}$ needs to be given by a theory
which in turn is presumably calibrated by other observables.

The third and final case, with $\eta\ll1$, is trivial.
Here the effects of kinetic
mixing are small either because $\chi$ itself is very small, or because
the hidden-sector $D_b$-term is much less than the intrinsic SUSY-breaking
scale itself.
An interesting example with $\langle D_b\rangle\sim\mz$ is provided by models
in
which SUSY-breaking is communicated to the visible world by a
$U(1)_X$ whose anomalies are cancelled via the Green-Schwarz
mechanism~\cite{anomalous}.
In such models, kinetic mixing in the $D$-terms provides only a small
correction to the masses of the MSSM particles unless $\chi\sim\CO(1)$.
(We will argue in the next section that such large values of $\chi$ are indeed
possible.)
Also note that one must still verify that loops of $U(1)_X$ gauginos coupled
to the MSSM particles do not
induce large masses proportional to the $F$-components in the hidden sector.

How small must $\chi$ be in the generic case in order to avoid either
destabilizing a model, or
at least providing large corrections to it (\ie, to avoid $\eta\gsim 1$)?
As an example,
let us consider two standard cases in which we again identify $U(1)_a$ with
the $U(1)$ of hypercharge:  that
of SUGRA-mediated SUSY-breaking in which we assume
$\langle D_b\rangle \sim\Lambda^2\sim\mz\mpl$, and that of gauge-mediated (GM)
SUSY-breaking in which we assume $\langle D_b\rangle \sim(\al_Y/4\pi)^{-2}
\mz^2\sim(100\tev)^2$. In these two cases, stability of the weak scale puts
upper bounds on $\chi$:
\beq
|\chi|~\lsim~\left\{\begin{array}{clc}(\mz/\mpl)/g_Y & \sim 10^{-16}
& \quad\quad{\rm(SUGRA)}\\
(\al_Y/4\pi)^2/g_Y & \sim 10^{-6}
& \quad\quad{\rm(GM)~.}\end{array}\right.
\label{bounds}
\eeq
Since $\chi$ is dimensionless and no new symmetries necessarily arise as
$\chi\to0$,
such small values of $\chi$ are unnatural
from the point of view of 't Hooft
and require either some conspiracy or some new symmetry of the theory.
This second possibility will be further explored in the next section.

\setcounter{footnote}{0}
\section{Kinetic Mixing in Field Theory and String Theory}

Having argued that non-zero $\chi$ can lead to large corrections to
the soft scalar masses, we now consider the typical
magnitude of $\chi$ which one would expect to be generated
both in field theory and in string theory.

\subsection{Expected magnitude of kinetic mixing in field theory}

\begin{figure}[t]
\centering
\epsfxsize=5in
\hspace*{0in}
\epsffile{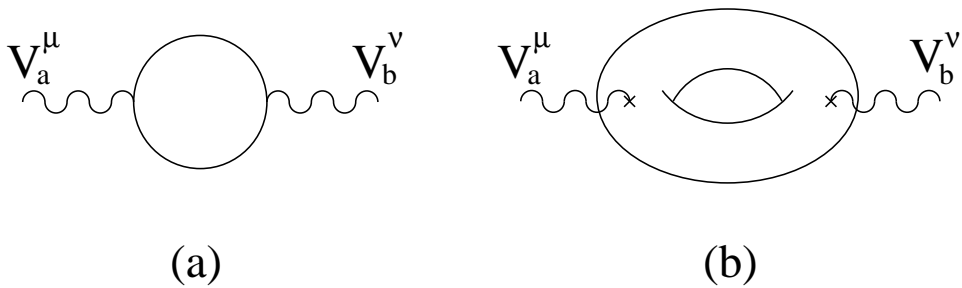}
\caption{(a) The one-loop diagram which
contributes to kinetic mixing in field theory, and
(b) its generalization to string theory.}
\label{fig:loop}
\end{figure}

Let us first consider the generation of $\chi$ in an effective field theory
context.
Kinetic mixing can be generated at one loop when there exist states which are
simultaneously charged under both $U(1)$ gauge factors. Consider a chiral
superfield
of mass $m$ with charges $Q_a$ and $Q_b$ under the two $U(1)$ gauge factors.
Such a chiral superfield contributes to the two-point polarization diagram
of Fig.~\ref{fig:loop}:
\beq
\Pi_{ab}^{\mu\nu}(\mu)~=~\frac{g_ag_b}{16\pi^2}\,Q_aQ_b\,
      \log\left(\frac{m^2}{\mu^2}
\right)
\left[k^\mu k^\nu - k^2 g^{\mu\nu}\right]~.
\label{eq:chiloop}
\eeq
In the effective Lagrangian,
this generates the operator $\frac{1}{2}F_{\mu\nu}^{(a)}F^{(b)\mu\nu}$
with coefficient~\cite{holdom}
\beq
\chi(\mu)~=~-\frac{g_ag_b}{16\pi^2}\,Q_aQ_b\,
      \log\left(\frac{m^2}{\mu^2}\right)~.
\label{eq:chiloop2}
\eeq
In general, at scales $\mu$ far from $m$, it is necessary to resum
the large logarithms using the renormalization group equations (RGE's):
\bea
 {dg_a\over dt}&=&\frac{1}{16\pi^2}\,g_a^3 b_{aa} \nonumber\\
 {dg_b\over dt}&=&\frac{1}{16\pi^2}\,g_b\biggl( g_b^2 b_{bb} -
                    2g_a g_b \chi b_{ab} \biggr)  \nonumber \\
 {d\chi\over dt}&=&\frac{1}{16\pi^2}\,\biggl(g_a^2 \chi b_{aa} +
               g_b^2\chi b_{bb} - 2g_a g_b b_{ab}\biggr)~
\label{eq:rge}
\eea
where $b_{ab}\equiv Q_aQ_b$ for the single superfield. For two or more
superfields, we take $b_{ab}\equiv \sum_i Q_a^{(i)}Q_b^{(i)}$.
Note that these RGE's hold only in the limit $\chi\ll 1$;
expressions valid for all $\chi$ can be found in Refs.~\cite{lepto,delag}.
As shown in Ref.~\cite{lepto}, large values of $\chi$ (\ie,
$0.1\lsim \chi\lsim 1$) can quite easily be generated by such
renormalization group running in realistic models,
thereby changing the low-energy phenomenology substantially.

%%%%%%%%%%%%%%%%%%%%%%%%%%%%%%%%%%%%%%%%%%%%%%%%%%%%%%%%%%%%%%%%%
%%%%%%%%%%%%%%%%%%%%%%%%%%%%%%%%%%%%%%%%%%%%%%%%%%%%%%%%%%%%%%%%%%%

In hidden-sector models, $Q_aQ_b$ is by definition zero for the states
below the SUSY-breaking scale. Thus, at one loop, non-zero values of $\chi$ are
not generated in the effective field theory below this scale.
Nonetheless it is entirely possible that non-zero $\chi$ can be
generated as a threshold effect
in the full theory at scales far above the SUSY-breaking scale,
a region about which one has in principle very little knowledge.
In the low-energy theory, such a value would not be suppressed by powers
of the high mass scale.

In order to estimate the typical size of such an effect, let us
consider a toy model with
two chiral superfields of charges $(Q_a,Q_b)$ and $(Q_a,-Q_b)$ and masses
$m$ and $m'$ respectively. Their joint contribution to $\chi$ has the form:
\beq
\chi~=~-\frac{g_ag_b}{16\pi^2}\,Q_aQ_b\,\log\left(\frac{m^2}{m'^2}\right)~.
\label{eq:toymodel}
\eeq
At scales far below $m$ and $m'$, this constitutes a threshold correction.
For $m-m'\sim m$, charges of ${\cal O}(1)$, and gauge couplings in the range
$1/60\simeq\al_Y(\mz)\lsim\{\al_a,\al_b\}
\lsim\al_Y(M_{\rm GUT})\simeq1/25$,
this leads to a contribution to $\chi$ of typical magnitude
\beq
    10^{-2}~\lsim ~\chi ~\lsim~ 10^{-3}~.
\label{eq:threshmag}
\eeq
This is only mildly dependent on the mass scale of the states through
the running of the couplings.
Given the bounds in Eq.~(\ref{bounds}), this clearly constitutes a
large effect.

One may wonder if it is possible to prevent the appearance of kinetic
mixing by requiring some property of the spectrum. Indeed there are two
possibilities.

Clearly, if one or both of the $U(1)$ gauge factors sits within an unbroken
non-abelian gauge symmetry, then kinetic mixing is not possible. However,
one could suppose that the spectrum of matter states fills out unsplit
non-abelian multiplets despite having broken off a $U(1)$ gauge factor from the
full
non-abelian gauge symmetry. While such a spectrum would indeed provide for
an exact cancellation of kinetic mixing because of the tracelessness of the
$U(1)$ generators on the non-abelian multiplets,
the presumed mass degeneracy of the states within such multiplets
is not stable against
radiative corrections and therefore non-zero kinetic mixing is generated after
all.  We will see in the next section that this is exactly the
situation in certain string models.

Another possibility is to forbid non-zero $\chi$ by
imposing a discrete symmetry. Such a symmetry would be a cousin of charge
conjugation which acts non-trivially on only the hidden $U(1)$.
In particular, consider embedding $U(1)_b$ at large scales into a non-abelian
group $G$ in which there exists a $G$-transformation $\Ga$ that inverts
$U(1)_b$.  Let us suppose that when $G$ breaks, it leaves,
in addition to $U(1)_b$, $\Ga$ as an unbroken discrete gauge symmetry.
In the low-energy theory, $\Ga$ acts on $U(1)_b$ (but not on the theory
as a whole) as a charge conjugation.
In this case the states charged under $U(1)_b$ appear in
degenerate conjugate pairs, and no $\chi$ can
be induced.  Equivalently, the kinetic mixing operator itself is forbidden
by the symmetry. It is worth noting that the types of symmetries which can
forbid the existence of a Fayet-Iliopoulos term are exactly of this form.
It is also very interesting
that such a symmetry necessarily implies the existence of
stable ``Alice strings'' with their delocalized charged
excitations~\cite{alice}.  Examples of such models include the $SU(6)\times
U(1)$ SUSY-breaking model in Ref.~\cite{dnns} and the
visible-sector left-right model of Ref.~\cite{span}.

Apart from this one exception, we will now argue that the ``ultimate''
high-scale theory, namely string theory,
naturally leads to a generation of $\chi\neq0$.

\subsection{Kinetic mixing in string theory: General formalism}

Since realistic string models often
have gauge groups containing many $U(1)$ gauge factors as well as
infinite towers of massive (Planck-scale) string states,
such theories serve as an ideal laboratory for estimating the magnitude
of kinetic mixing effects.
First, we shall show how kinetic mixing effects can be calculated
in string theory.  Then, we shall discuss the typical sizes that
string theory predicts for such effects.

We begin, however, with few preliminary remarks concerning generic
properties of string spectra.  Unlike the case in field theory,
the tree-level string spectrum consists of states
whose masses are quantized in units of the Planck mass.
Only the states of lowest energy (the massless states)
are observable. Furthermore, in string theory, the freedom
to alter or redefine the charges of states is extremely limited,
for one must satisfy a number of worldsheet consistency constraints
stemming from worldsheet conformal invariance, modular invariance,
and worldsheet supersymmetry.
One therefore seeks to construct a particular string theory
(or ``string model'') which satisfies all of these constraints
and which is also ``realistic'' ---  {\it i.e.}\/,
whose massless states are those of the MSSM or its extensions
coupled to $N=1$ supergravity.
Indeed, once a particular string model
is constructed that satisfies all of these constraints,
its entire spectrum of massless and massive
string states is completely fixed, and their
quantum numbers cannot be altered.
Thus string theories provide a very rigid structure
in which certain phenomenological properties such as kinetic
mixing may be meaningfully tested.

With this in mind,
our goal is to repeat the one-loop field-theoretic
calculation discussed above, only now in a string-theoretic context.
Certain parts of this calculation are straightforward.
As illustrated in Fig.~\ref{fig:loop},
the string generalization of the field-theoretic vacuum
polarization diagram is a torus amplitude with two vertex operators
inserted on the worldsheet (corresponding to the gauge bosons
of the $U(1)_a$ and $U(1)_b$ gauge factors).
All of the string states (both massless and massive)
propagate in the torus and thereby contribute to this one-loop diagram.
However, in order to evaluate this diagram in a fully consistent
way in string theory, we must perform this calculation in a
non-trivial background which satisfies the string equations of
motion and which, in particular, takes into account not only the contributions
from the gauge interactions but also their gravitational back-reaction
in the presence of a suitable regulator \cite{Kaplunovsky,Kiritsis}.
Indeed, {\it a priori}, both gauge and gravitational terms
contribute in generating a non-zero value of $\chi$ in the string-derived
low-energy effective Lagrangian ${\cal L}$ in Eq.~(\ref{firstequation}).

Such string calculations have previously been performed in
the case that the two inserted gauge bosons come from the {\it same}\/
$U(1)_a$ gauge factor;  in such cases the result describes the coefficient
of the corresponding $F^{(a)}_{\mu\nu}F^{(a)\mu\nu}$ factor in ${\cal L}$,
and is relevant for the study of the so-called ``heavy string threshold
contributions'' to the corresponding gauge coupling $g_a$
\cite{Kaplunovsky,Kiritsis,review}.
Recall that in such cases, the effect of
this one-loop torus amplitude is to correct the
gauge coupling $g_a(\mu)$ at a scale $\mu$ according to
\beq
     {1\over{g_a^2(\mu)}}~=~k_a\,\frac{1}{g_X^2}~+~
     {b_a\over 16 \pi^2}\,\ln{ M_X^2\over{\mu}^2}~+~{1\over 16 \pi^2}
\,\Delta_a~.
\label{onelooprunning}
\eeq
Here $M_X$ and $g_X$ are the unification scale and coupling
(here to be identified with the string scale and coupling), while
$k_a$ is the $U(1)_a$ normalization factor.  This normalization factor
is analogous to the string
Ka\v c-Moody level which appears for non-abelian gauge groups \cite{DFM},
and is defined analogously as the coefficient of the double-pole term in
the worldsheet operator-product expansion (OPE) of the $U(1)_a$ current $J_a$
with itself:
\beq
   J_a(z)\,J_a(w) ~=~ {k_a/2\over (z-w)^2 }~+~ {\rm regular}~.
\label{OPEone}
\eeq
For example, in the case of hypercharge $U(1)_Y$,
the MSSM predicts $k_Y=5/3$.
Finally, the threshold correction $\Delta_a$ in Eq.~(\ref{onelooprunning}) is
defined as
\beq
  \Delta_a ~\equiv ~ k_a \,{\cal Y}~+~   \int_{\cal F}
     {d^2\tau\over {\rm Im}\,\tau}\,
     \biggl\lbrack B_a(\tau) ~-~ b_a \biggr\rbrack~.
\label{deltadef}
\eeq
Here ${\cal Y}$ is a so-called ``universal term'' which receives contributions
from both gravitational back-reaction
and universal gauge oscillators  \cite{Yterm,Kiritsis};
the integral over the complex parameter $\tau$ in the modular region
${\cal F}$ represents a summation over conformally inequivalent tori;
$b_a$ is the one-loop beta-function
coefficient (calculated by considering only the massless string spectrum);
and $B_a(\tau)$ represents a supertrace over {\it all}\/ string states
with arbitrary left- and right-moving spacetime masses $M_L$ and $M_R$:
\beq
        B_a(\tau) ~\equiv~
        {\rm Str} ~\left\lbrace \left({1\over 12}-\overline{Q_H}^2\right)\,
               {Q_a}^2~\,
         \overline{q}^{\alpha' M_R^2 }\,q^{\alpha' M_L^2}\right\rbrace~,~~~~~
                 q\equiv e^{2\pi i\tau}~.
\label{Bdef}
\eeq
Here $Q_a$ is the gauge charge operator for the gauge group factor
$U(1)_a$, $\alpha'$ is the Regge slope,
and $\overline{Q_H}$  is the spacetime helicity operator.
In this context, recall that in field theory, the one-loop $\beta$-function
coefficient can be written as
$b_a= {\rm Str}\, \lbrack(1/12-h^2) Q_a^2\rbrack$
where $h$ is the helicity operator (with $h=0$ for scalars,
$\pm 1/2$ for fermions, and $\pm 1$ for vectors).
Likewise, in Eq.~(\ref{Bdef}), $\overline{Q_H}$ is the analogous
helicity operator for states of arbitrarily high spin.
Since we are working in the context of string theories with
spacetime supersymmetry, we can omit the factor of $1/12$ in Eq.~(\ref{Bdef})
since its contribution is proportional to ${\rm Str}\, {\bf 1}=0$.

Given these results, it is relatively straightforward to generalize
the calculation to the case when the two vertex-operator insertions
correspond to {\it different}\/ $U(1)$ gauge factors.
Taking $\chi=0$ at the unification scale at tree level, one then finds
that the string spectrum naturally generates a non-vanishing
value for $\chi$ at one loop given by
\beq
       \left\lbrace\frac{\chi}{g_ag_b}\right\rbrace(\mu) ~=~
	{b_{ab}\over 16\pi^2}
       \,\ln{ M_X^2\over{\mu}^2}
      ~+~{1\over 16 \pi^2} \,\Delta_{ab}~.
\label{RGEchi}
\eeq
Note that according to the RGE's in Eq.~(\ref{eq:rge}), it
is the combination $\chi/g_ag_b$ which runs analogously to the usual
gauge couplings $1/g_a^2$.
Thus we see that Eq.~(\ref{RGEchi}) is indeed
the string analogue of the field-theoretic
expression given in Eq.~(\ref{eq:chiloop2}).
The first term in Eq.~(\ref{RGEchi}) represents the contribution
from the massless string states,
with $b_{ab}$ serving as their ``mixed'' beta-function coefficient
\beq
        b_{ab} ~=~ -\,{\rm Str}_{\rm massless}~  \overline{Q_H}^2\, Q_a\,Q_b~,
\eeq
while the second term in Eq.~(\ref{RGEchi})
is the kinetic mixing threshold correction due to the
infinite tower of massive string states.
This threshold correction is defined as
\beq
      \Delta_{ab}~=~
         k_{ab}{\cal Y}~+~ \int_{\cal F} {d^2\tau\over {\rm Im}\,\tau}\,
     \biggl\lbrack B_{ab}(\tau) ~-~ \,b_{ab} \biggr\rbrack~.
\label{Deltaabdef}
\eeq
Here ${\cal Y}$ is the same universal term that appears in the
above gauge coupling calculation, and $k_{ab}$
is defined analogously to Eq.~(\ref{OPEone}) via
the OPE between the $U(1)$ currents $J_a$ and $J_b$:
\beq
     J_a(z) \, J_b(w)~=~ {k_{ab}/2\over (z-w)^2} ~+~ {\rm regular}~.
\label{OPEtwo}
\eeq
Note that $k_{aa}=k_a$.
Likewise, $B_{ab}(\tau)$ is the kinetic mixing supertrace
\beq
        B_{ab}(\tau) ~\equiv~
        -\,{\rm Str}  ~\overline{Q_H}^2\, Q_a\,Q_b~\,
         \overline{q}^{\alpha' M_R^2 }\,q^{\alpha' M_L^2}~.
\label{Babdef}
\eeq
Thus, since $\Delta_{ab}$
contributes as a threshold correction due to purely massive string states,
it is this object which is our primary focus.

As indicated in Eq.~(\ref{Deltaabdef}),
$\Delta_{ab}$ generally contains two
separate contributions.   However, while the second term
in Eq.~(\ref{Deltaabdef}) is highly model-dependent,
the first term $k_{ab}{\cal Y}$ is universal and thus can be
evaluated in a general setting.
Indeed, even though ${\cal Y}$ generally receives contributions
from both gravitational back-reaction and gauge oscillators,
calculating such contributions is unnecessary
in the present case because taking $\chi=0$  at tree level
is tantamount to requiring that $U(1)_a$ and $U(1)_b$
be orthogonal gauge gauge factors in our string model.  This
in turn implies that OPE's between their currents should vanish,
or that $k_{ab}=0$.
This can also be seen by
writing our two $U(1)$ gauge factors
as linear combinations of the
22 elementary $U(1)$ worldsheet currents $J_i$ ($i=1,...,22$) of the heterotic
string,
\beq
            U(1)_a~=~ \sum_{i=1}^{22}  \, a_i^{(a)}\, J_i~,~~~~~
            U(1)_b~=~ \sum_{i=1}^{22}  \, a_i^{(b)}\, J_i~,
\eeq
for it is then
straightforward to show that
the one-loop universal contributions are
all proportional to
\beq
           k_{ab}~\equiv~ 2 \, \sum_i a_i^{(a)} a_i^{(b)}~.
\eeq
Indeed,
since it is known \cite{DFM}
that $k_a= 2\sum_i [a_i^{(a)}]^2$,
we see that once again we have the relation $k_{aa}=k_a$.
However, orthogonality of the $U(1)_a$ and $U(1)_b$ gauge factors
explicitly means that $\vec a^{(a)} \cdot \vec a^{(b)}=0$.
This again implies that $k_{ab}=0$.

Thus, we conclude that
only the second term in Eq.~(\ref{Deltaabdef})
can contribute to one-loop kinetic mixing effects in string theory.

\subsection{Kinetic mixing in string theory:  Expected magnitude}

Given these results, we now wish to determine the expected
size of the kinetic mixing parameter $\chi$ in string theory.
In other words, given two orthogonal $U(1)$ gauge factors in a given string
model,
how large a mixing contribution $\Delta_{ab}$ will typically be generated
by summing over the contributions of all of the massive string states?

In order to estimate the size of such effects,
one can calculate $\Delta_{ab}$ for
different orthogonal $U(1)$ gauge factors
in a variety of four-dimensional string models.
In general, for string models with moduli of order one,
it turns out that one typically obtains
quite sizable results, with $\Delta_{ab}$ lying in the range
\beq
             10^{-1} ~\lsim ~|\Delta_{ab}|~\lsim ~10^{+1}~;
\label{results}
\eeq
moreover, in generic models with larger moduli,
$|\Delta_{ab}|$ increases dramatically.
We stress that the results in Eq.~(\ref{results}) indeed record the
contributions from only the {\it massive}\/ string states, for any
undesired kinetic mixings due to massless string states are
explicitly removed from the integrand of Eq.~(\ref{Deltaabdef})
through the subtraction
of the mixed beta-function coefficient $b_{ab}$.
Thus, {\it a priori}\/, Eq.~(\ref{results}) sets the scale
for the expected contributions
to kinetic mixing from the massive states in string theory.
This in turn implies that at the weak scale, we have
\beq
     3g_b\times10^{-4} ~\lsim~ \chi(\mz) ~\lsim~ 3g_b\times10^{-2}~.
\label{results2}
\eeq
Here we have interpreted $U(1)_a$ as hypercharge, and
have used Eq.~(\ref{RGEchi}) to translate from $\De_{ab}$ to $\chi$.
As can be seen from comparison with Eq.~(\ref{bounds}),
this is a very large effect which has the potential
to destabilize the supersymmetric gauge hierarchy.

However, one possible objection to this result is the fact
that it is not sufficient to demand mere orthogonality of
the different $U(1)$ gauge factors.  Indeed, we are
interested in the case when these $U(1)$ gauge factors are
also presumably {\it hidden}\/ from each other.
Of course, if two $U(1)$ gauge factors are truly hidden from each
other in this way, then we have $Q_a Q_b=0$ for each state in
the string spectrum.   This in turn implies
that all of the above terms vanish, and that no kinetic mixing
is generated at one loop.
However, it turns out that imposing this condition is far too severe
in the context of string theory --- we should really only require
that the {\it massless}\/ ({\it i.e.}\/, observable) string states
satisfy $Q_a Q_b=0$. Indeed, given a particular choice of
charge assignments for the massless states, the string self-consistency
constraints do not generally permit all massive string states to
satisfy $Q_aQ_b=0$.
In such cases we would then have $b_{ab}=0$, but $\chi$ would still receive
contributions from $\Delta_{ab}$ as in Eqs.~(\ref{RGEchi}) and
(\ref{Deltaabdef}).

Unfortunately, no realistic string models have been constructed
for which $Q_a Q_b=0$ for all states in the massless spectrum.
Indeed, in string theory, it has become traditional to refer to
a gauge group as ``hidden'' merely if it does not couple to
those states in the massless spectrum which correspond to
the Standard Model particles ---
such hidden gauge symmetries
can nevertheless couple to the extra states beyond the Standard Model
which also appear in the massless spectrum and which have
Standard Model quantum numbers.
We will therefore refer to such groups as being only ``semi-hidden''.
Thus, we see that in general, semi-hidden $U(1)$ gauge group factors
can still give rise to mixed coefficients $b_{ab}\not= 0$.
As indicated in the first term of Eq.~(\ref{RGEchi}),
this may serve as an additional string-theoretic source
for kinetic mixing.  To be conservative, however, in what follows
we shall focus only on the remaining terms,
namely the corrections $\Delta_{ab}$ due to
the {\it massive}\/ string states.

In order to calculate $\Delta_{ab}$ for the
case when $U(1)_a$ and $U(1)_b$
are semi-hidden relative to each other, we have evaluated
Eq.~(\ref{Deltaabdef}) within the context of
various special semi-realistic string models
which have been constructed in the literature \cite{ALR,278,274}.
All of these string models have $N=1$ supersymmetry, three chiral
generations, and phenomenologically interesting gauge groups
such as $SU(3)\times SU(2)\times U(1)_Y$ and $SO(6)\times SO(4)$.
Moreover, they also give rise to semi-hidden gauge symmetries
including extra $U(1)$ gauge factors.
We thus seek to calculate the mixing that takes place in these
models between the hypercharge $U(1)_Y$
and one of these extra semi-hidden $U(1)$ gauge factors.
Indeed, such a calculation is completely analogous to the
gauge-coupling threshold correction calculations that were performed for
these string models in Ref.~\cite{DF}.

The result we obtain, however, is quite surprising:
in each of those realistic string models, it turns out that
$\Delta_{ab}$ in Eq.~(\ref{Deltaabdef}) vanishes exactly!
This occurs {\it in spite}\/ of the fact that these models contain
states (even massless states) which simultaneously carry both hypercharge
and semi-hidden $U(1)$  charge.  Indeed, even though such states
exist in these models, their contributions nevertheless cancel
level-by-level across the entire massless and massive string spectra.

It is easy to see why such cancellations arise
in these particular models.
In the $SO(6)\times SO(4)$ string model \cite{ALR}, for example,
this cancellation arises due to the field-theoretic GUT mechanism
discussed previously:  the hypercharge $U(1)$ generator is
embedded within the larger non-abelian group structure $SO(6)\times
SO(4)$, and therefore the trace over all multiplets
cancels exactly (or equivalently, the kinetic mixing
term in the effective Lagrangian would not be invariant
under the full non-abelian
$SO(6)\times SO(4)$ gauge symmetry, and thus cannot exist).
A similar phenomenon arises even in those
string models \cite{278,274} whose low-energy gauge groups
are already broken down to $SU(3)\times SU(2)\times U(1)_Y$
at the Planck scale.

However, this cancellation is ultimately unstable against
a variety of effects beyond tree level because the exact cancellation
of $\De_{ab}$ relies on the exact tree-level degeneracy of the states
at each mass level.
Effects that can destroy this exact degeneracy include GUT symmetry breaking,
Dine-Seiberg-Witten shifts of the string ground state (which
are generally necessary in order to break anomalous $U(1)$ gauge symmetries
and restore spacetime supersymmetry \cite{DSW}),
and low-energy supersymmetry breaking.
In fact, even if these effects do not destroy the degeneracy,
renormalization group flow down to low energies inevitably will.
For example, in the string model of Ref.~\cite{278},
the contribution to kinetic mixing from extra massless electroweak
doublet states beyond the MSSM is cancelled by those from extra Standard Model
singlets and color triplets.
While it is noteworthy that such cancellations arise in the first place,
they are clearly unstable against radiative corrections.
Thus, since such cancellations are not protected by any symmetries,
they must be viewed as accidental artifacts of the tree-level spectrum.

Given this situation, we would like to calculate the amount of kinetic mixing
that will be induced in these models by all of these effects.
 {\it A priori}, we would need to calculate the exact spectrum of massless
and massive string states after all of the above effects have been included.
We would then use this complete spectrum as an input into
Eq.~(\ref{Deltaabdef}), thereby iteratively calculating the true value
of $\chi$. Unfortunately, such a calculation is beyond present capabilities.

However, one can estimate the mass splittings generated in the string
spectrum by each of the above effects. To be conservative, we will
estimate the contributions from the splittings of only the {\it massive}
states;
it is clear that splittings of the massless states would yield a very large
effect.
Our procedure will be as follows.  For simplicity, we assume
that the contributions of the states in the tree-level spectrum cancel pairwise
in $\De_{ab}$.  We can then use the results in Eq.~(\ref{eq:toymodel}),
together with an estimate of the splitting $\De m$ between $m$ and $m'$,
to determine the resulting contribution to $\chi$ from each pair.
Each of the effects which splits the pairs has a characteristic scale
associated with it.  Mass splittings due to GUT symmetry breaking are typically
$\De m\sim 10^{16}\gev$.
In the case of vacuum shifting, while it is true that at least one
scalar field obtains a vacuum expectation value of approximately $10^{17}\gev$,
the effects on the remaining states are typically suppressed
due to discrete symmetries which forbid low-dimension terms in the
superpotential.  Indeed, a typical mass scale for such splittings has been
argued~\cite{alon}\ to be as low as $10^{11}\gev\lsim\De m\lsim10^{14}\gev$.
Next, in the case of splittings induced by SUSY-breaking, the appropriate scale
is the fundamental scale of SUSY-breaking in the hidden sector, since by
definition
the states are charged under the $U(1)_b$ which couples directly to the
SUSY-breaking sector. In the supergravity case we have $\De m\sim10^{11}
\gev$, while in the gauge-mediated case we have $\De m\sim 10^4\gev$.  Finally,
in all three of the above cases, RGE flows generally produce additional
splittings
which will be of the same order as their respective splitting scales $\De m$.

The above $\De m$ splittings then produce the following estimates for
$\chi$:
\beq
\frac{|\chi|}{g_b C}~\sim~\left\{\begin{array}{cc} 10^{-4}
& \quad\quad{\rm(GUT)}\\
10^{-6}\,-\,10^{-9} & \quad\quad{\rm (vacuum~shift)} \\
10^{-9} & \quad\quad{\rm (SUGRA)} \\
10^{-16} & \quad\quad{\rm (GM)~.} \end{array} \right.
\label{results3}
\eeq
In the above estimates, $g_b$ is the hidden-sector $U(1)_b$ coupling, while
the quantity $C$ parametrizes the effect of the summation over all newly
split pairs of states throughout the string spectrum.  As such, we estimate
the combined effect of such a summation to be approximately
\beq
         10^1~\lsim~ C ~\lsim~ 10^2~.
\eeq
We emphasize again that in this estimate, we are being conservative
by ignoring the effects of the splittings of massless string states
which can also be sizable.
We also point out that the effects in Eq.~(\ref{results3}) are not
exclusive of each other;  a given string model will typically be
subject to a simultaneous combination of these effects.

Thus, our conclusions regarding the size of
kinetic mixing in string theory are as follows.
In general, the expected scale for kinetic mixing effects
is given by Eq.~(\ref{results2}).  However, in certain semi-realistic
string models, there is an accidental cancellation of kinetic mixing
effects in the tree-level spectrum.  In such cases,
non-zero kinetic mixing will then be generated
by the effects considered above, and the resulting magnitudes for kinetic
mixing
from each effect are given in Eq.~(\ref{results3}).
We see from these results that there exists considerable variation in
the possible amounts of kinetic mixing.  Nevertheless, as
can be seen by comparing these results with those of Eq.~(\ref{bounds}),
they are all in the interesting range for low-energy phenomenology.

\setcounter{footnote}{0}
\section{Conclusions}\label{sec:conc}

We have demonstrated the existence of an important new mechanism for the
communication of supersymmetry breaking from a hidden sector
to the visible sector. This mechanism applies when there
exists in the hidden sector a $U(1)$ gauge symmetry that is not
isolated from SUSY-breaking, and it relies on a $D$-term interaction which is
induced by the supersymmetric generalization of kinetic mixing.

The new contributions to the soft squared masses for scalars are
proportional to hypercharge $Y$.
In cases where the $D$-term of the hidden $U(1)$ is larger than the
weak scale (in particular, of order the scale of fundamental SUSY-breaking),
this leads to phenomenologically
disastrous consequences unless the mixing parameter $\chi$
is very small. In particular, one must have
$|\chi|\lsim 10^{-6}$ in the case of gauge-mediated models
with $D$-terms $\sim (100\tev)^2$, and $|\chi|\lsim 10^{-16}$ in the
case of supergravity-mediated models with $D$-terms
$\sim (10^{10}\gev)^2$.

We have argued that since $\chi$ is a
renormalizable interaction, its value is sensitive to physics at
all mass scales. In particular, substantial values of $\chi$ can be generated
at arbitrarily high mass scales in both field theory and
string theory contexts.  We have shown how to calculate
the amount of kinetic mixing generated in string theory,
and found that estimates of the magnitude of $\chi$ in both
field theory and string theory often lead
to values in excess of the above limits.

Thus we conclude that the kinetic mixing
parameter $\chi$ should be considered as a very natural additional
measurable parameter describing the soft SUSY-breaking spectrum.

%=========================================================================

\bigskip
\bigskip
\leftline{\large\bf Acknowledgments}
\bigskip
We are pleased to thank J.~Bagger, E.~Kiritsis, S.~Sethi,
F.~Wilczek and especially K.S.~Babu for discussions.
This work was supported in part by DOE Grant No.\ DE-FG-0290ER40542
and by the W.M.~Keck Foundation.

%=========================================================================
\bigskip
\bigskip

\bibliographystyle{unsrt}

\end{document}